\documentstyle[12pt]{article}

\begin{document}
\textwidth 170 mm
\textheight 240 mm
\topmargin - 0.8 cm
\oddsidemargin -0.2cm
\evensidemargin -0.2cm
\headheight 0pt
\headsep 0pt
\topskip 9mm

\newcommand{\beq}{\begin{equation}}
\newcommand{\eeq}{\end{equation}}
\newcommand{\bq}{\begin{quotation}}
\newcommand{\eq}{\end{quotation}}
\newcommand{\BFACE}[1] {\mbox{\boldmath $#1$} }

\def\theequation{\arabic{section}.\arabic{equation}}

\title{Seeking inspiration from the Standard Model in order to go
beyond it}

\author{
{\sc Holger Bech Nielsen\thanks{This manuscript has
                                evolved appreciably since a
                                talk was given with the
                                above title.}
                                \& Christian Surlykke} \\
{\sl The Niels Bohr Institute,} \\
{\sl Blegdamsvej 17, 2100 K\o benhavn \O, Denmark } \\
\and
{\sc Svend Erik Rugh\thanks{On leave from The
                  Niels Bohr Institute (1/1 - 1/8 1993). E-mail addresses:
                  ``HBECH@nbivax.nbi.dk'', ``SURLYKKE@nbivax.nbi.dk'',
                  ``RUGH@mitlns.mit.edu'' and
                  ``RUGH@nbivax.nbi.dk''}   } \\
{\sl Center for Theoretical Physics,} \\
{\sl Massachusetts Institute of Technology,} \\
{\sl 77 Massachusetts Avenue,} \\
{\sl Cambridge, MA 02139, U.S.A.} \\
}
\date{}
\maketitle

\setcounter{footnote}{0}

\begin{center}
{\small{\bf Abstract}}
\end{center}

{\small We look at various features of the Standard Model
with the purpose of
exploring some possibilities of how to seek
physical laws beyond it, i.e.
at even smaller distances.
Only parameters and structure which are not calculable
from the Standard Model is considered useful
information. Ca. $90$ bits of information contained in the
system of representations in the Standard Model are explained
by four reasonable postulates.
A crude estimate is that there is of the order of
$\sim 2 \times 10^2$ useful bits of unexplained
information left today.
There are several signs of the fact that the Standard Model is
a low energy tail of a more fundamental theory (not yet
known).
However, some worries are expressed as concerns how far
the exploration of the physics
beyond the Standard Model can proceed - if we
are to be inspired from
these $\sim 2 \times 10^2$ bits alone.

}

\newpage

\setcounter{page}{2}

\section{Introduction}

It is a wonderful fact, and we should not forget to appreciate
it, that the collaborative activities here
on our planet - which take place roughly $\sim 10^{17}$ seconds
after the ``big bang''  -
have been able to design experiments
creating physical conditions which are substantially
removed from the natural energy scales on earth.
So far we have been able to set up experiments which
can give information about
the extreme physical conditions in the universe
back to approximately $\sim 10^{-10}$
seconds after big bang!

Moreover, a set of (mathematical) regularities
which account for the observed behavior
of Nature under these extreme conditions
(with energies of the order of $\sim 10^{11}$ eV per
microscopic degree of freedom)
have been found and are compactly
described in the generally accepted
Standard Model of the electroweak and strong interactions.

Questioning \cite{Weinberg} whether we will be able
to identify a set of principles which captures how
Nature operates at even earlier
times, i.e. at even higher energy,
it is worthwhile to speculate
how to be guided towards such principles?

The situation is today that theoreticians (the creators  of the
Glashow Salam Weinberg Ward
model and QCD, i.e. of the Standard Model) have caught up with
experiment to such an extent that making further theoretical progress is up to
the problem that there is very little information about the laws of Nature
which is not
- in principle - explained by this Standard Model. So we have to make
careful use of the little information yet left
unexplained.\footnote{By unexplained we here mean
{\em unexplainable} in terms of predictions from
the Standard Model itself, i.e.
we do not count the huge amount of hadronic data, say, which we today
think are explainable entirely within the Standard Model itself, though
requiring non-perturbative techniques with are beyond present abilities.}

\section{The Standard Model (input and output) }

The Standard Model provides not only detailed perturbative
- and even some nonperturbative results - of specific
processes but explains also many general features such as
symmetries. For example we may indeed consider the following
features derivable from the Standard Model:

\begin{enumerate}
\item Parity, charge conjugation, and time-reversal symmetry
for strong and electromagnetic interactions.\footnote{Provided, though,
that  the theta-term coefficient $\Theta_{QCD}$ of
QCD is very small.}

\item Chiral symmetry and hereunder approximate flavour-symmetries (due to the
smallness of some quark masses), e.g. isospin and Gellman-SU(3)
symmetry for strong interactions. Conservation of baryon B and
lepton number L and especially B-L are derivable with a high accuracy
(at low temperature).

\item Custodial symmetry \footnote{Custodial symmetry is the extra
$SU(2)$-symmetry of the  Higgs-field physics arising from the fact that the
Higgs-field potential $V(|\phi|^2)$ and the Higgs-field kinetic term are
invariant under $O(4)\sim SU(2)\times SU(2)$ symmetry - considering the
two-component Higgs-field 4 real components  transforming as a four-vector
under
$O(4)$ - while it is only the one of these $SU(2)$'s which is identified with
the Glashow-Salam-Weinberg (GSW) $SU(2)$ gauge group.
The remaining  $SU(2)$ is the
custodial symmetry group.}
\end{enumerate}

\noindent
Besides
many symmetries, a lot of
parameters in e.g. atomic or nuclear physics
are derivable from the
Standard Model\footnote{Note, that often
{only very little} of the Standard Model
is used in the derivation
of such parameters in atomic or nuclear physics.
Remember that the ``Standard Model'' has absorbed in it
quantum mechanics, Q.E.D., the Maxwell equations etc.
so we can consider e.g. Bohr's formula to be a consequence of it.}
(think e.g. of a
parameter like Rydbergs constant $ Ry = m_e e^4/ 2 \hbar^{2}$
in atomic physics or the decay rates for super-allowed transitions in nuclei).

However, some features
must be imposed at the outset. These features rather defines
the Standard Model - they cannot be derived from it. We
could call them ``input'' into the Standard Model.
Some of these features concerns symmetries, e.g.:

\begin{enumerate}
\item 3+1 dimensional Poincar\'{e} invariance.
\item Gauge symmetry under the gauge Lie algebra $U(1)\times SU(2)\times
SU(3)$ or, better, Lie {\em group} $S(U(2)\times U(3))$.
\item The fermion Weyl particle representations with its repetition
in three generations (families), cf. table \ref{representationer}.
\item The Higgs or Higgs-replacement is doublet with y/2=1/2.
\end{enumerate}

\noindent
Not so transparently related to symmetries  are
the general principles of quantum mechanics and
quantum field theory (and renormalizability) which is also to
be considered input structure into the Standard Model.
In addition to these input principles there are a number of unexplained
parameters:
\begin{quotation}
\noindent
3 lepton masses, 6 quark masses, 3 gauge coupling constants,
the expectation value of the Higgs field $<\Phi>$,
1 topological angle (close to zero), 4 quark mixing angles, i.e.
all in all of the order of 19 external (unexplained, but already measured)
parameters or so.
\end{quotation}

There are also a few not yet measured parameters which are only accessible
through cosmology --- if at all --- or through better accelerators: The Higgs
mass, the weak topological angle, $\Theta_{SU(2)}$, and to some extent even the
top mass.

{\small
\begin{table}
\label{representationer}
\begin{tabular}{|l|c|c|c|}
\hline
  \multicolumn{4}{|c|}{\em Representation of Standard Model}\\
  \multicolumn{4}{|c|}{\em Matter Fields}\\
\hline
  \begin{tabular}{l}
    $U(1)\!\times\!SU(2)\!\times\!SU(3)$ \\
    representation \\
    \hspace{4 mm}$(y/2,I_W,\underline{a})=$
  \end{tabular} &
  \begin{tabular}{c}
    $1^{st}$\\
    family
  \end{tabular} &
  \begin{tabular}{c}
    $2^{nd}$\\
    family
  \end{tabular} &
  \begin{tabular}{c}
    $3^{rd}$\\
    family
  \end{tabular}\\
\hline
  \begin{tabular}{l}
     $(1/6,1/2,\underline{3})$:\\
     \hspace{4 mm} $q_L \stackrel{CP}{\sim} (\bar{q})_R$
  \end{tabular} &
  $\left(
    \begin{array}{ccc}
      u_r& u_y& u_b \\
      d_r^c & d_y^c & d_b^c
    \end{array}
   \right )_L$  &
  $\left(
    \begin{array}{ccc}
      c_r& c_y& c_b \\
      s_r^c & s_y^c & s_b^c
    \end{array}
   \right)_L$  &
  $\left(
    \begin{array}{ccc}
      t_r& t_y& t_b \\
      b_r^c & b_y^c & b_b^c
    \end{array}
   \right)_L$ \\
\hline
  \begin{tabular}{l}
    $(-2/3,0,\underline{\overline{3}})$: \\
    \hspace{4 mm} $(\bar{u})_L \stackrel{CP}{\sim} u_R$
  \end{tabular} &
  $(\bar{u}_r,\bar{u}_y,\bar{u}_b)_L$ &
  $(\bar{c}_r,\bar{c}_y,\bar{c}_b)_L$ &
  $(\bar{t}_r,\bar{t}_y,\bar{t}_b)_L$ \\
\hline
  \begin{tabular}{l}
    $(1/3,0,\underline{\bar{3}})$: \\
    \hspace{4 mm} $(\bar{d})_L \stackrel{CP}{\sim} d_R$
  \end{tabular}&
  $(\bar{d}_r,\bar{d}_y,\bar{d}_b)_L$ &
  $(\bar{s}_r,\bar{s}_y,\bar{s}_b)_L$ &
  $(\bar{b}_r,\bar{b}_y,\bar{b}_b)_L$ \\
\hline
  \begin{tabular}{l}
    $(-1/2,1/2,\underline{1})$: \\
    \hspace{4 mm}
    $\left(
       \begin{array}{c}
         \nu\\
         l
       \end{array}
     \right)
     \stackrel{CP}{\sim}
     \left(
       \begin{array}{c}
         \bar{\nu}\\
         \bar{ l}
       \end{array}
     \right)$
  \end{tabular}&
  $\left(
     \begin{array}{c}
       \nu_e \\
       e^-
     \end{array}
   \right)_L$ &
  $\left(
     \begin{array}{c}
       \nu_{\mu}\\
       \mu
     \end{array}
   \right )_L$ &
  $\left(
     \begin{array}{c}
       \nu_{\tau}\\
       \tau
     \end{array}
   \right)_L$ \\
\hline
  \begin{tabular}{l}
    $(1,0,\underline{1})$: \\
    \hspace{4 mm} $(l^+)_L \stackrel{CP}{\sim } l^-_R$
  \end{tabular} &
  $e^+$ &
  $\mu^+$ &
  $\tau^+$ \\
\hline
\hline
  $(-1/2,1/2,\underline{1})$: &
  \multicolumn{3}{c|}{Higgs?}\\
\hline
\end{tabular}
\caption[xxx]{
This table lists all the 15 irreducible representations
of left handed Weyl fields in the Standard Model, the representation structure
being denoted as an ordered triple in the left column. The first
item of this triple $(y/2, I_W, \underline{a})$
is the weak hypercharge in the normalization, often called
$y/2$, the next is the weak isospin $I_W$, and the third is the dimension of
the associated irreducible representation of the color $SU(3)$-group
with an underlining, the anti-triplet is denoted $\underline{\overline{3}}$.
The lower indices $r$, $y$ and $b$ refer to the three colors
``red'', ``yellow'' and ``blue''.
The upper index $c$ denotes that it is not exactly the flavour
indicated by the symbol
but the Cabibbo-rotated one which goes together with the
associated quark-flavour of the $2/3$-charged quark. The Higgs
field is written with ``?'' to denote that it may not exist.  \\
Charge-parity conjugation will give the representations, as indicated in the
table, through the relations:\\
$q_L \stackrel{CP}{\sim} (\bar{q})_R$,
$(\bar{u})_L \stackrel{CP}{\sim} u_R$,
$(\bar{d})_L \stackrel{CP}{\sim} d_R$,
$\left( \begin{array}{c}
         \nu\\
         l
       \end{array}  \right)
\stackrel{CP}{\sim}
\left( \begin{array}{c}
         \bar{\nu}\\
         \bar{ l}
       \end{array} \right)$ and
$(l^+)_L \stackrel{CP}{\sim } l^-_R$
}
\end{table}}

\subsection{Why and how to measure the information content in the Standard
Model ?}

How long can theoretical physicists go beyond the Standard Model without new
progress in experimental physics? This is measured by the amount of information
in the structure and parameters in the Standard Model not yet explained.
Any proposal for a theory beyond the Standard Model would namely
have to defend its truth by explaining some of this information.

But how shall we measure, more quantitatively, this information?
If we
want to teach contemporary particle physics to a layman or an alien, we will
explain that it is described by a gauge theory, what the gauge
group is, which representations is used and finally the value of the external
parameters.
But, prior to that, we will also
have to teach him general physics, and
the concepts necessary to define a gauge field theory etc. This process would
require a huge amount of information, or text-bits,
and such a number can not be of interest
for the purpose of estimating the
scientific value of a model in theoretical physics.\footnote{To bring the
concepts in theoretical physics, e.g. the concepts
defining the Standard Model, in contact with daily language (and ultimately
with elementary elements of perception) is a very complicated task.
The outcome of such an exercise will depend on the cultural context
in which we develop physics. Besides, if we were able to rewrite the
entire Standard Model into the form of an text-string consisting of
elements of language (words) in direct contact with elementary perceptual
experiences, the Standard Model would look so complicated that its
beauty and simplicity would be hidden.
Somewhat analogously, a simple and beautiful program written in Pascal
looks horribly complicated if written in terms of
``machine language''.}

Rather the tables below may be viewed as a kind of  ``shopping guide''
which the reader may find useful
when buying a ``model of everything''or a ``model behind''. The
scientific value  of such models is given,
then, by the number of bits they are able to explain. The price of a model
is, in this analogy, the number of bits required to define it.   \\

\noindent
{\small
As a terrifying example of explaining a presumably {\em negative} number
of bits one may mention the ``world machinery'' model described
in \cite{HolgerSvend1} set up to explain the 3
bits of the space-time dimension, using a rather involved set
of model assumptions,  representing a
number of bits which is complicated to evaluate, but -
most likely - is greater than 3.

As a less terrifying example we may mention some work by
S. Dimopoulos, L.J. Hall, S. Raby and A. Rasin \cite{Hall}.
In a scheme taking over a Gorgi-Glashow supersymmetrized $SU(5)$
G.U.T., or rather $SO(10)$ scheme, they postulate some operators
responsible for the mass matrices
and yield so remarkable agreements with
experimental masses and mixings
that it looks, at first, as being able to explain 30 bits! (they
explain 13 parameters
from 7 input parameters). \footnote{However,
by closer inspection it turns out
that about four bits of the information has been fitted into their scheme
by inserting zeroes in the mass matrices (strongly) inspired from the
already known experimental numbers. Moreover, a place where a
larger amount of adjustment could potentially have sneaked in is in
selecting the composite operators involving several successive
exchanges of $\underline{45}$'s and intermediate fermion propagators.
These ans{\"a}tze lead to Clebsch-Gordan-coefficient-like factors.
These factors are
though of order one and rational - but they may still represent
a fitting of discrete parameters.
A crude estimate shows that there will only be of the order
of 30-22-4= 4 bits left as the truly explained information, but even that
might be quite suggestive of at least some truth being there.
See \cite{Hall} for the more general features
really explaining the agreement. } }

\vspace{0.5 cm}

\noindent
Let us sketch how we may crudely define
- i.e. in a somewhat arbitrary way - the number
of ``bits'' needed in setting up a model as the Standard Model,
i.e. the amount of information contained in the input structure and input
parameters needed to specify the model.

We take this number of bits of unexplained information to be given
by
\begin{equation}
log_2 \; \left( \begin{array}{c}
                \mbox{Number of models with parameter-} \\
                \mbox{system which have similar} \\
                \mbox{or less amount of complexity}
                \end{array}
                \right)
\end{equation}

But what do we mean by a model which have a complexity comparable
to that of the Standard Model?
In fact, this is a very complicated question for which we will not
offer an exhaustive analysis here. It may not have an objective
answer (unaffected by some theoretical bias).

Premature considerations of the difficulty of
varying some (sub)structure and yet obtain a well
defined class of models (thus necessarily keeping other
structure fixed) may also be found in sec.3.3 in
\cite{HolgerSvend2}.

Here, we consider a very restricted class of models
- models which only deviate from the Standard Model in details\footnote{
In fact, we only vary the gauge
Lie algebra, the representations of it
and the space-time dimensionality.} -
and count {\em as separate models the same model with different values of
the parameters provided these parameters deviate more than
the present experimental uncertainty}. Talking about the models
of same or lower degree of complexity we really have in mind to
restrict ourselves to consider quantum field theories, so that
we just have to consider as possible those models which are obtained
by specifying a gauge Lie algebra (that can easily be a direct
product of several simple ones - as it is the case for the
Standard Model) and a system of representation for
particles/fields of various spins.

It is important - but somewhat of an {\em arbitrary} choice - that we
do not include more complicated representations
and Lie algebras nor larger parameter values than those appearing
in the Standard Model itself. The reason for this
(somewhat arbitrary)
convention in the definition of the number of unexplained
bits of information (to be listed in our tables) is that there would
otherwise be the problem that there are eventually infinitely many
Lie algebras, infinitely many possible - usually reducible -
representations and infinitely many possible parameter values
- also if one consider values deviating by less than the experimental
uncertainty as identical values
(so we really count the number
of uncertainty intervals). Even after the suggestion of this
principle for counting ``bits'' the exact implementation in the various
cases, Lie algebra, Lie group, system of representations,
dimensions etc. still involve a bit of {\em arbitrary} choices.
Especially, we shall include what we call ``order of magnitude information''
for those parameters for which
a priori (though theoretically prejudiced)
expectations exists (usually from
dimensional arguments).  \\

\noindent
{\bf Distance measure to a ``theory of everything'' } \\
A ``theory of everything'' ($T.O.E.$), with the word
``everything'' taken in an elementary strict sense, contains
{\em zero bits} of unexplained
information ! One could in principle build up a distance measure,
counting (a lower estimate of) the number of unexplained ``bits''.
How far is a given model $\; T \; $ from being a $T.O.E.$ ?
\begin{equation}
|| T - T.O.E. || = \left( \begin{array}{c}
                      \mbox{\# of unexplained} \\
                      \mbox{bits of information}
                      \end{array}
                      \right)
\end{equation}
We shall see that the Standard Model is at least $\sim 2 \times 10^2$ bits away
in distance (a number which grows
with the increasingly precise measurements)
from a ``Theory of Everything''!

We remark, that the number of bits we are able to arrive
at are {\em lower estimates}, both because of our
conventions of not counting models which are more complicated than
the Standard Model, and because we are
anyway not able to take into account the
possible models (simpler or more complex)
that we are not able to even think about.

\subsection{Amounts of unexplained information in the
Standard Model (our definitions)}
For a quantity - a parameter - $k$ for which there is no special expectation
w.r.t. order of magnitude, and which is (presently)
measured with an uncertainty
$\bigtriangleup  k$
we define the
amount of (unexplained) information contained in such a parameter (yet
to be explained) to be\footnote{If there were really no quantity of the same
dimension to compare with the value would be useless for making theories, but
we imagine using $c=G=\hbar=1$ to define our unit system;  cf. footnote in
section \ref{gravity}.}
\begin{equation}
\log_2\frac{k}{\bigtriangleup k}.
\end{equation}
If there is an order of magnitude expectation - usually on grounds
of dimensional arguments - we suggest to take into account the information
in measuring the order of magnitude  of the ratio of the quantity $k$ to
the expectation called $scale$ (which is only defined up to a factor
$e = 2.7..$, say), by imagining that the numerical value
of the logarithm $| \ln (k/scale)| $ following our convention
could have been smaller but not larger in a ``simpler model''. Since
$scale$ is
only defined order-of-magnitudewise it can not be significant if $k$
by has the same value as $scale$ to better than, say, a factor
$e$.
The
total amount of information in the measured value of the parameter $k$
is therefore taken to be
\begin{eqnarray} \label{totalinf}
\log_2 \frac{\max\left\{1,|\ln \frac{k}{scale}|\right\}}
                      {\bigtriangleup \ln k}
&\simeq & \log_2\frac{\max\left\{1,|\ln\frac{k}{scale}|\right\}}
                  {\bigtriangleup k/k}\\
& = & \log_2 \frac{k}{\bigtriangleup k } +
             \max\left\{0,\log_2|\ln\frac{k}{scale}|\right\}.
\end{eqnarray}
The second term in this expression is what we call
``the order of magnitude information''
\begin{equation} \label{orderinf}
        \left( \begin{array}{c}
        \mbox{order of magnitude} \\
        \mbox{ information }
        \end{array}
        \right)
        = \max\left\{0,\log_2 |\ln \frac{k}{scale}|\right\}.
\end{equation}
When several similar quantities such as the
quark masses all deviate in the same direction (by being smaller say)
from the expected $scale$,
the Higgs expectation value in vacuum $<\phi>_{vac}$,
we shall presumably rather refer these parameters to each other and only one
of them to the a priori $scale$ $ <\phi>_{vac}$ ; {I.e. our}
expected $scale$ for a quark mass goes down once
we know that another quark (or lepton) is surprisingly light.
By this convention we minimize the amount of order of magnitude information
to be listed in our table.

Consider, as an example, the parameter $\Theta_{QCD}$
for which it is only known
that  $\Theta_{QCD} < 10^{-9}$.
It must be counted as having an experimental uncertainty equal to
the value so there
is only order of magnitude information in the smallness of
$\Theta_{QCD}$. The expected value (since it is an angle between 0 and
$2 \pi$) is of the order
of $\Theta_{QCD} \sim 1$. Therefore there
is $\sim \log_2 \ln (10^9) \sim
4 $ bits of order of magnitude information (which is also a substantial
amount of order of magnitude information
carried by only one parameter).\\

\noindent
{\bf Standard Model - parameter information (cf. table 2)} \\
For the weak scale
$<\phi>_{vac}$ itself
($<\phi>_{vac} \sim 246 \; GeV $)
one often takes the expected $scale$ to be
the Planck energy scale with the motivation of the prejudice that this
is the ``most fundamental'' scale to use.\footnote{However, there
may very well be new
physics at much lower scales, for example, say, of the order
of $\sim 10^2-10^3 TeV$, from which $< \phi >_{vac}$ arises.}
Even for dimensionless quantities such as the finestructure constants
one should
look for some ``natural'' unit to find the $scale$ to be used.
However, since we take
double logarithm ``$\log_2\ln$'' in (\ref{orderinf})
it does not matter much what we use
for the $scale$, say the typical critical finestructure constant
$\alpha \sim 1/20$ or just simply $\alpha \sim 1$.

\begin{table}
\begin{tabular}{||c||c|c|c||} \hline
\multicolumn{4}{||c||}{{\em Standard Model - parameter
information}}\\ \hline \hline
&
\begin{tabular}{c}
$log_2 (k/\Delta k)$ \\
information\\
\end{tabular} &
\begin{tabular}{c}
Order of \\
magnitude \\
information \\
$\log_2 |\ln (k/\mbox{scale})|$ \\
\end{tabular} &
Total \\ \hline \hline
Weak scale $<\phi>$ or $G_F$ &  15 & \mbox{0 to 5}$^{*}$
&  \mbox{15 to 20} \\ \hline
Fine structure constants &  31 &  \mbox{0 to 3} &  \mbox{31 to 34} \\ \hline
Charged lepton masses    &  46 &   \mbox{7 to 9} &  \mbox{53 to 55} \\ \hline
dsb-quark masses         &   6 &   \mbox{6 to 8} &  \mbox{12 to 14} \\ \hline
uct-quark masses         &   5 &   5 &  10 \\ \hline
\begin{tabular}{c}
CKM-matrix \\
including CP-violation \\
\end{tabular}            &  13 &   5 &  18 \\ \hline
$\theta$-strong          &   0 &   4 &   4  \\ \hline
\begin{tabular}{c}
In total for \\
Standard Model \\
true parameters \\
\end{tabular}            & 116 &  \mbox{27 to 39} & \mbox{143 to 155} \\ \hline
\end{tabular}
\caption[XYZ]{{\small Estimated information in the parameters of
the Standard Model. The $*$ for the 5 bits
of order of magnitude information for the weak scale $< \phi > $
indicate that it is an upper estimate which corresponds to the
{commonly asserted} - though debatable - theoretical prejudice that
the Planck scale, or G.U.T. scale say, is the natural expected scale of
``any phenomenon'' - and thus also of the weak scale.}
}
\end{table}

Because of  confinement the very concept of a quark mass is not so clean.
One must distinguish the ``constituent quark mass'' defined as being the
one used in the nonrelativistic quark model(s) and the ``current algebra
quark mass'' - which is the one most directly connected to the
Yukawa couplings. The latter is determined from the breaking of the
chiral symmetry caused by this current algebra mass and reflects itself
as a nonzero mass to the approximate Nambu-Goldstone
bosons $\pi$, $K$ and $\eta$. It is the current algebra masses or,
rather, the related Yukawa couplings which are to be thought of
as parameters, ``input parameters'', of the Standard Model. Because of the
uncertainties in the techniques of extracting these numbers (= current algebra
masses) from QCD it is presumably fair to say that they are not
determined better than to say three significant digits on base two
or say one significant digit on base ten ($2^3 \sim  10$). For the heavy
quarks similar uncertainties appear
although
the technique of estimating the masses is somewhat different from that
for the light quarks.   \\

\noindent
{\bf Standard Model - discrete information (cf. table 3)} \\
Table 3 shows the discrete information in those
input features
of the structure of the Standard Model
for which alternatives are easy to imagine.
({For example} we do not
count the possibilities of exchanging the principles of
quantum mechanics with something else).

The estimated amount of information in the gauge Lie
algebra expresses
that there are of the order of $\sim 2^6$
Lie algebras with rank less than or equal to that of
the Standard Model ($S(U_2 \times U_3) \simeq U(1) \otimes SU(2) \otimes
SU(3)$ has rank four).

\begin{table}
\begin{tabular}{||c|c|c||} \hline
\multicolumn{3}{|c|}{{\em Standard Model -
discrete information} }\\ \hline
&
\begin{tabular}{c}
Before understanding \\
Weyl representations \\
\end{tabular} &
\begin{tabular}{c}
After understanding \\
Weyl representations \\
\end{tabular} \\ \hline
Gauge Lie$
\left\{
\begin{array}{c}
\mbox{algebra} \\
\mbox{group} \\
\end{array}
\right.$  &
\begin{tabular}{c}
6 \\
- \\
\end{tabular} &
\begin{tabular}{c}
- \\
8 \\
\end{tabular} \\ \hline
Weyl representations        &  92 &  0 \\ \hline
Higgs representation       &  6 &  4 \\ \hline
$3+1$ dimensions           &  3 &  3 \\ \hline
Spin distribution          &  8 &  6 \\ \hline
Total                      &  115 &  21 \\ \hline
\end{tabular}
\caption[xxx]{{\small Listing of information
for various types for discrete settings
in the structure of the Standard Model. The first column with numbers
shows the information unexplained before
we introduce the four assumptions listed in table 4 (section 3.2 below).
The last column gives the unexplained numbers after this
explanation has been taken into account.
} }
\end{table}

The item ``Higgs representation'' contains  $\log_2$ of an estimate of
the number of representations smaller than or equal to the actual
Higgs representation. The dimension and signature 3+1 is one
possibility between the possibilities
0+0, 1+0, 2+0, 1+1, 3+0, 2+1, 4+0, 3+1, 2+2, which makes up 9
possibilities (using our convention of not counting larger numbers than
those in the Standard Model) and thus they contain
$\log_2 9 \approx 3$ $bits$.
The item ``Spin distribution'' is supposed to count the information
in how many {fields (or particle types) we have with} different spin.
In this
connection we take into account that a massless photon or Yang Mills particle
is so different in spin structure from a massive spin 1 that it should
be counted as a different possibility. In this item is
also included the
information that there are 3 generations.
How we arrive at the item ``Weyl representations'' is best explained in
connection with the table 4, but really the main point is to
count $\log_2$ of the number of ways 15 irreducible representations
can choose between the $13\cdot 2\cdot 3 = 78$ possibilities for having
$y/2=-1,-5/6,-2/3,-1/2,-1/3,-1/6,0,1/6,1/3,1/2,2/3,5/6,1$; weak isospin
$I_W=0,1/2$; and the color $SU(3)$ representation being
$\underline{1}$,$\underline{3}$ or
$\underline{\overline{3}}$. ($15\cdot \log_2 78 =93 \; bits$).
One of the $93$ $bits$
corresponds to the
knowledge of weak interactions coupling to {\em left} handed
quarks and leptons rather than to {\em right} handed quarks and leptons,
and that {particular} bit is therefore not meaningful without
an a priori standard (convention) - which do not exist -
for what is left and right.

\subsection{Some remarks about the very little amount of
unexplained information in gravitation theory}
\label{gravity}
{We restrict attention to the Standard Model and - essentially - }
ignore gravity which
is the other branch of laws of physics which is well established today.
Compared to the Standard Model the situation with respect to the number of
parameters and structure measured in gravity is that there is
strictly speaking only one properly measured parameter -
the Newtonian gravitational constant - and one parameter you would
expect to be able to measure with present accuracy,
the cosmological constant $\Lambda$. In Planck units,
the expectation $\Lambda \sim 1$ (i.e. $scale \sim 1 \; m_{Pl}^4$)
would be the natural scale for the cosmological constant (and such values
are achieved in many quantum gravity models).
However, the measured value is less than $10^{-120} m_{Pl}^4$.
{Thus}, if we extended our counting of unexplained bits
to include gravity, the cosmological constant would - in much analogy
to $\Theta_{QCD}$ - represent no unexplained genuine measurement
information since its measurement uncertainty is
still as large as the quantity itself, but a lot of unexplained
order of magnitude information $\sim \log_2 \ln (10^{120}) =
\log_2 (120 \cdot 2.3) = \log_2 276 \sim 8 \; bits$ corresponding to
the cosmological constant problem: Why is the cosmological constant
so exceedingly small compared to a priori expectations ?

One might formally count $\log_2 (G/ \bigtriangleup G) \sim 13 \; bits$
as the content in the Newtonian gravitational constant, but
from a model building point of view one must accept
some dimensionful quantities
- just to set the scale\footnote{
Explicitly, we need the measurement of three dimensionful
quantities in order to set a standard by which we measure for example
$length$, $time$ and $mass$ (and it is then a profound fact -
the validity of this fact requires a bit of reflection -
that the dimensions of all other quantities in physics, e.g.
the quoted $\sim 19$ parameters of the
Standard Model, may be expressed in terms of these three
standard units).
The three dimensionful quantities are often selected to be
the Newtonian gravitational coupling constant $G$, the
velocity of light $c$ and the Planck constant $\hbar$.
Thus, we can {\em not} include the measurements of $G$, $c$ and $\hbar$
- in our table 2 - as useful parameter information
since the function of these parameters, by choice, merely is to inject
a scale into the entire system of parameters of the Standard Model.
}. The measurement of {such quantities}
is then not really useful in
model building except in putting the scale for the other
quantities. Having used the gravitational constant for defining the unit
system we cannot also count it as carrying information.

In total we thus essentially only get the 8 bits of the
cosmological constant puzzle from gravity yet to be explained. The
theory of gravity is also structurally already so elegant that very
little information is needed to specify the model and presumably only of
the order of a couple of discrete information bits are as
yet unexplained: The graviton is spin 2 but could say a priori have had
less spin and perhaps have had all helicities.
In all we thus have of the
order of $\sim 8 + 2 = 10$  bits more than described above by including
gravity.

{We note, that since} there are so few bits to explain {\em within}
the theory of gravity itself
the scientific value of some {\em ``quantum gravity''} theory
almost unavoidably has to be measured by the amount of predictive power it has
concerning data which do {\em not}
deal with gravitation theory itself,
i.e. it
{\em has to predict data which lie outside of
gravitation theory !}
For example, if the ``quantum theory of gravitation'' is able to predict
some of the 19 unexplained parameters of the Standard Model
(of the electroweak and strong interactions) then the ``quantum gravity''
theory {may} be justified {(and gain some belief)}
this way.

\subsection{Information from cosmology? (thermal equilibrium phases
are strong ``cosmological filters'')}

{Including cosmology in} the considerations of seeking beyond
the Standard Model we may have some more data at our disposal -
even with todays accelerators and measurements. We have the Hubble
expansion rate (extremely crudely) and the background radiation
temperature - the famous $3^{0}$ K radiation, really $2.75^{0}$ K -
and some primordial abundances. Most informative are presumably possible
correlations and fluctuations over space of say the temperature
(COBE results) and the distribution and
motions of galaxies.


In cosmology we really see the
big bang physics through an extremely cloudy
{\em filter!} It is to be
hoped - of course - that this may turn out not {completely} to
be the case - in view of the forseeable limitations
in {\em our} abilities of putting
substantially more than present energies $\sim 100 GeV$ on a
single elementary particle.

It appear to us, that if {\em thermal equilibrium phases} are reached at
certain
stages in the evolution of the universe \footnote{
Thermal equilibrium phases are
reached in several
evolutionary stages of the early universe.
For example, thermal equilibrium phases
in the formation of quark-gluon plasma at energy scales $\sim \; GeV$
are, presently, contemplated to be
reached within $\sim \; fm/c \sim 10^{-23} \; seconds$
whereas the corresponding phase in the Universe has a time span
of the order of $\sim 10^{-5} \; seconds$. So there is a lot of time
to form thermal equilibrium phases of quark-gluon plasma during the
cooling of the universe!

Also, the plasma of charged electrons and charged nuclei has
a lot of time to reach thermal equilibrium w.r.t. {\em atomic physics}
degrees of
freedom before the transition from plasma to a gas of neutral atoms
takes place (approximately $\sim 10^5 \; years$ after big bang). In addition to
conserved quantum numbers such as baryon number, lepton numbers and electric
charge only the number of each type of primordially produced nuclei
survives this  equilibrium.}
then only very little information
of the physics that went on {\em before} that phase can reach us today.
It is  difficult for a signal to survive through an equilibrium phase.
Basically only {\em conserved}
{quantum numbers survive, like baryon and lepton
numbers and energy.} If the thermodynamic equilibrium is
not reached globally, there may survive some information in the
correlations or, rather, in the spatial variations in the densities of
these conserved quantities (and the temperature).

\section{Which information in the Standard Model is most inspiring?}

It is of interest to identify (in retrospect) those features
in the disciplines of classical mechanics (macroscopic bodies),
solid state physics, atomic physics and
nuclear physics, say, which point towards
some regularities at {the energy level
above} (cf. the quantum staircase) and try that way
to develop intuition and more ``general principles'' for how
to search for underlying structure.

An inspiring way to start is presumably to
identify those features which are not
yet understood, and among them, single out
the most surprising ones (``mysteries'') as the most inspiring.
Two forms of surprises are:

(1) {\em Paradoxes }

(2) {\em Dimensional surprises }

\vspace{0.5 cm}

\noindent
{\bf (1) Paradoxes:} A true
paradox has a potential of giving an important
hint that something is wrong somewhere.
The attempt to cure such a paradox is therefore a good starting point
for guessing intelligently some underlying structure.
Cf. the paradox {(in the end of last century)} of the ultraviolet
catastrophe in black body radiation
(leading to the Planck hypothesis) and the paradox
of why the orbits in the atom do not fall into the center, leading
to Niels Bohrs hypothesis of ``stable quantum states''
(by which Bohr in the same go killed around $\sim 4$ bits by
explaining the Rydberg constant\footnote{This
relation has been confirmed by the
later decimals { arriving due to the better}
measurements since Niels Bohrs
times. { So today this relation}
explains more than 4 bits. Due to relativistic
corrections $\sim \alpha^2$ the simple Bohr relation should however
not be true to {better than $\sim 10^{-4}$ corresponding}
to $\sim 13$ bits of predictive precision.}
as the combination $m_e e^4/ 2 \hbar^{2}$).

As an example of a paradox, existing today, one could perhaps
point to the non-regularizability of the chiral Weyl particles
in the Standard Model (c.f. discussion and
references in \cite{HolgerSvend1,HolgerNinomiya}).
Another paradox might be that it is most likely
not possible to make the cut off
go to infinity (non-perturbatively) in electrodynamics.
Quantum Electro Dynamics is not renormalizable at the non-perturbative level,
in the sense that according to \cite{Rakow} one can find various physical
quantities varying with the cutoff for given renormalized coupling and mass.
Since the Standard Model contains a $U(1)$ group (with similar
properties as electrodynamics)
this paradox is very likely also shared by
the Standard Model.\footnote{
A rescue may be to embed the Standard Model in some
grand unified theory in which asymptotic freedom is ensured, although the
presence of (a) Higgs field(s) may continue to present a problem. }

\vspace{0.5 cm}

\noindent
{\bf (2) Dimensional surprises:}  If a given phenomenon in Nature occurs
with a strength (for example, with a transition amplitude in
quantum mechanics)
which is much smaller - or, perhaps, even vanishes -
relative to what we expect on dimensional
grounds it signals interesting underlying principles.

Two well known examples (by now well understood) illustrates this
point:

$\bullet$ The vanishing of the monopole mode (which is the first
one would expect to dominate
on purely dimensional grounds with no understanding
of the underlying physics) in the electromagnetic dipole radiation
and the vanishing of both
the monopole and dipole modes (the first contributing mode is the quadrupole)
in gravitational radiation carry signals of the
symmetry structure of electromagnetism and
general relativity
(and are explained e.g. in terms of the helicity structure of the
respective force carriers, the photon and the graviton).

$\bullet$ An example of a strongly suppressed transition amplitude
(relative to what we should expect from dimensional analysis)
which involve more detailed knowledge of
the structure of the Standard Model is
the smallness (but verified existence) of
{\em neutral flavour changing currents}. Such currents
have a suppression factor
$$ \sim \sin \Theta \cos \Theta \; (\frac{m_c^2 - m_u^2}{m_W^2}) $$
due to the clever cancellations between the generations (GIM-mechanism)
in the Standard Model.\footnote{In fact, it is very hard
to {\em compete} with the
Standard Model - when it comes down to constructing an underlying model
which gives the same suppression factor.
For example, ``technicolor models'' have
difficulties in reproducing this suppression factor.}

``Neutral flavor changing currents'' is a
wonderful example of getting inspiration from a feature
which is surprisingly suppressed relative to expectations from
dimensional arguments. In the sixties, when only
three quarks $u,d,s$ were known,
there was no mechanism (known) to suppress the neutral flavour changing
current. {It guided} Glashow (and Bj\"{o}rken) to predict a new
quark --- $c$.

\subsection{Unsolved problems and ``mysteries'' -
Which are the most important ? }

While the above examples are by now explained within the Standard
Model itself, let us now proceed to identify similar - but totally
unexplained - ``puzzles'' or ``mysteries'' of the Standard Model which
may very well give valuable hints of structure
(e.g. symmetries) {\em beyond} the Standard Model.

Some of the input parameters of the Standard Model
have in fact quite
strange values (they are examples of what we call
``dimensional surprises'' and they carry
correspondingly a large amount of ``order of magnitude information''
in table 2): \\

\noindent
$\bullet$ $\Theta_{QCD}$ is zero
with great accuracy
($\Theta_{QCD} < 10^{-9}$)
while it ``ought''
to be of order of $\pi$ or so, since rotational angles are in the compact set
$[0, 2 \pi]$. Note that the corresponding angle
for the $SU(2)$ group, $\Theta_{SU(2)}$, is not measured. \\

\noindent
$\bullet$ The smallness of quark and lepton masses relative to
the weak interaction scale $< \phi >_{vac}$ or the Fermi constant $G_F$, say.
This could point to some approximative symmetries (and associated
conserved quantum numbers), cf. sec.3.4.  \\

\noindent
$\bullet$
The large generation gaps in the fermion masses (i.e. presumably
in the Higgs Yukawa couplings).  \\

\noindent
$\bullet$ Why is the
Higgs-scale (the scale of weak interactions say) so exceedingly
low compared to the Planck-scale or to some grand unification
scale, say, in case such a unification should exist ?  \\

\noindent
$\bullet$ If we include
gravity, why is the cosmological constant $\Lambda$
so exceedingly small (or zero)?   \\

\noindent
As an example of a mystery a widely announced example is
the interpretation of quantum mechanics. \\

Let us now turn to some examples - mainly our own - in
which we claim to look fairly unprejudiced (straightly)
at
the data  - i.e. the parameters and
structure of the Standard Model itself.

\subsection{First observation: Only little information is
stored in the representation system of the fermions}

It is to be expected that it is easier to extract inspiration from
the structural information {(discrete information)} than from the
numbers (the parameters), which need more
model building to become inspiring, and among the
former the information
in the Weyl fermion representations which {contain} around
90 $bits$ seems promising to attack at first glance.
However, we shall see that with a few rather reasonable assumptions,
listed in table 4,  we can explain essentially all these 90 $bits$\footnote{
Note, that the very large number of bits, $\sim 90$ $bits$,
contained in table 4, is due to the large number
(45) of particles and {the fact that each particle
could in principle have a different representation.}
If we impose the constraint that we have a system of three
generations with identical representations the above number of bits
would reduce roughly by a factor of 3.

It should be emphasized, also, that some
of our estimates of the number of $bits$ in table 4 are somewhat crude.}:

\begin{table}
\begin{tabular}{||c|c|c|c||c|c||} \hline
\multicolumn{6}{||c||}{{\em Information in the
representation system} }\\ \hline \hline
Small             &  Anomaly     &  Mass       &  Charge  &
Information &  Excluding \\
represen-   &  constraints &  protection &  quantization &
left        &  handedness \\
tations  & & & & & \\ \hline \hline
$\bullet$ &           &           &            & 93   & 92    \\ \hline
$\bullet$ & $\bullet$ &           &            & 58   & 57    \\ \hline
$\bullet$ &           & $\bullet$ &            & 91   & 90    \\ \hline
$\bullet$ & $\bullet$ & $\bullet$ &            & 56   & 55    \\ \hline
$\bullet$ &           &           & $\bullet$  & 55   & 54    \\ \hline
$\bullet$ & $\bullet$ &           & $\bullet$  & 19   & 19    \\ \hline
$\bullet$ &           & $\bullet$ & $\bullet$  & 45   & 44    \\ \hline
$\bullet$ & $\bullet$ & $\bullet$ & $\bullet$  &  1   &  0    \\ \hline \hline
\end{tabular}
\caption[XYZ]{{\small Table showing how various combinations of the
four assumptions mentioned in the text reduce the amount of
information  contained in the representation system
(cf. table 1) of the matter
fields (fermions) in the Standard Model.
The numbers represent the information content measured in bits
which is not explained by the combination
denoted by the $\bullet$'s.
The true numbers are in the right handed column of numbers (as described
in the text).
} }
\end{table}

$\bullet$ {\bf Mass protection.}
One observation, {seen from table 1}, is the constraint of ``mass
protection'' which implements the constraint that under
no circumstances do we find both an irreducible representation and
its charge conjugate. That is, when we for example
find left handed dsb-antiquarks with
the representation $(y/2=1/3,I_W=0,
\underline{a} = \underline{\overline{3}})$ then we do
not find the charge conjugate representation
$(y/2=-1/3,I_W=0,\underline{a} = \underline{{3}})$.
{Now, the crucial observation is to note, that}
these two representations
allow together a mass term
$$ \sim m \psi_{\bar{r}}^T {\cal C}^T \psi_{r} + h.c.$$
that does not make use of the Higgs breaking of the
gauge group.
At more fundamental scales there may very well be particles which
are {\em not} mass protected. Such particles would have a mass
corresponding to the typical mass scale of the theory at the
more fundamental scale. This could be of the order of some
hundred $TeV$ (or even
of the order of the Planck scale, we might speculate).
The fact, that the Standard Model
obeys this ``mass protection rule'' tells us presumably that
it is a {\em low energy tail} of whatever is the more fundamental theory.
It only contains those particles that are protected from getting
``normal masses'' (from the point of view of
the theory beyond). From
the point of view of the Standard Model and from {an}
experimental point of
view, however, these ``normal'' masses are huge!
\par
$\bullet$ {\bf Charge quantization.}
Another ``observation'', which we {may} extract from table 1, is
a fact which we simply have put in for
{phenomenological} reasons\footnote{This
fact, which is the charge quantization rule, was known long before the
Standard Model was constructed and was rather implemented
- than predicted - in the construction of the Standard Model. }:
All particles in the Standard Model has to obey a charge quantization
rule which is a generalization of the well known Millikan charge
quantization rule,
\begin{equation}
Q=-t/3 \; (mod \; 1)\;\;\;\;\mbox{or equivalently}\;\;\;\;
y/2+d/2+t/3=0 \; (mod \;1)
\end{equation}
where $y/2$ {is the weak hypercharge,}
$d$ denotes ``duality'' which means $d=1$ for weak isospin $I_W$ being
half-integer, and $d=0$ for $I_W$ integer, while $t$ is triality for the
color representation meaning that $t$ modulo 3 counts the number of
triplet representations $\underline{3}$ needed to build up a representation
from which the representation in question can be extracted. E.g.
triality is $1$ for the triplet $\underline{3}$ and $0$ for
the singlet $\underline{1}$, whereas for
the anti-triplet $\underline{\overline{3}}$
it is $ -1 = 2 \; (mod \; 3)$.
\par
O'Raifeartaigh and L. Michel \cite{FroggattNielsen}
have argued that {the information in this}
quantization rule
can be packed into the requirement that the representations of the
Standard Model be representations (truly and not only ray-representations)
of the {\em group}
$ SMG  =
S(U(2)\times U(3))
\; (\subset SU(5)) $ which is specified as
$$
\left.\left\{\left( \begin{array}{ccccc}
u_{11}&u_{12}&0&0&0\\u_{21}&u_{22}&0&0&0\\
0&0&v_{11}&v_{12}&v_{13}\\0&0&v_{21}&v_{22}&v_{23}\\0&0&v_{31}&v_{32}&v_{33}
\end{array} \right) \right| u\in U(2)\;, \;  v \in U(3)\; and \;\det u
\cdot \det v =1 \right\}.
$$
The information of the charge quantization is thus compactly
packed (with use of fewer unexplained bits)
by saying that this is the Standard Model {\em group}, instead of just the
Lie
algebra. Since there are naturally more groups than Lie algebras up to a
given rank or a given dimension of the group, there are a couple of more bits
of information in knowing the group than the algebra; 8 for the group while
only 6 for the algebra, say. The $8 - 6 = 2 $ bits invested
pay off by explaining
around 37 to 55 bits (cf. table 3). So using the group is a strong scientific
progress compared to using the algebra.
\par
$\bullet$ {\bf Anomaly constraints.}
Even if a quantum field theory {at the classical level seems to be} gauge
invariant, there can be anomalies quantum mechanically.
{This means that it
is indeed not gauge invariant: It has gauge anomalies!} An intuitive
{idea about what happens} when one has anomalies may be
achieved by thinking of
the anomaly resulting from pumping Weyl particles up from the Dirac sea,
which has infinitely many particles, so the Dirac sea can remain
intact even after
pumping up some particles\footnote{ This is the effect of the infinite
hotel: the hotel is infinite - it has infinitely many rooms - but
it is full. Now there come one guest more, and indeed he can find place:
he gets number 1, the guest there is then moved to room number 2, and
guest in room number 2 is moved to 3, and so on. In the infinite hotel
this procedure works satisfactorily.}. In this way it seems that
some particles have been produced, or if pumped down have been destroyed.
If the particle(s) in question carry gauge charge such an effect will
spoil gauge invariance. This thread to gauge invariance is circumvented
in the Standard Model by a very sophisticated cancellation between the
particles of different types being pumped up and down: their charges are
chosen so cleverly that the net gauge charge being
pumped up (minus the amount pumped down) is just zero for all the
types of gauge charge under all the possible field configurations.
In four dimensional space time the anomalies arise in perturbation theory
from triangle diagrams with Weyl-particles going around the triangle loop.
Gauge bosons are attached to
the three corners of the triangle.
For each choice of the three gauge bosons
{which couple} via such a triangle
diagram the contribution to the anomaly from the full system of
Weyl fermions in the model must cancel in order that it shall not develop
true breaking of the gauge symmetry. A model as the Standard Model
will only remain consistently gauge invariant at the quantum level
provided these cancellations take place, and that imposes severe restrictions
on the system of Weyl representations. Imposing these necessary conditions
for the consistency thus reduces the number of allowed Weyl representation
systems and thus the amount of bit information yet to be explained.
There {are} always trivial solutions obtained by imposing parity or
charge conjugation invariance, but that would be totally forbidden by the
requirement of ``mass protection''. Under the anomaly requirements needed
for consistency is also the Witten discrete anomaly and the mixed anomaly.
The former says that there must be an even number of weak isospin doublets
($I_W=1/2$), and the latter comes from the requirement that the interaction
with the gravitational field must not cause gauge symmetry violation.
{It implies} that the summation over all the Weyl particles in the model
of the weak hypercharge $y/2$ should be zero.

The anomaly constraints - which mathematically are expressed as
6 different algebraic constraints - are indeed a very powerful set of
constraints. Note from table 4 that removing the anomaly constraints
from the Standard Model makes it possible to construct of the
order of $\sim 2^{45}$ models (instead of one).

\par
$\bullet$ {\bf Small representations.}
With the
three assumptions above concerning
the system of Weyl particle representations there are actually no
solutions with lower representations than those in
the ordinary generations in the
Standard Model.
In this sense we can claim that the
Weyl fermion representations
in the Standard Model
are remarkably small! By the assumption of small representations in the
table we simply mean that we only have allowed
representations of ``size''
(dimension and or charge) up to that
occurring in the Standard Model.
Since we - {anyway} - by our convention only consider the small
representations in our counting of bits,
we actually do not relax this assumption when we compute
the number of bits yet to be explained.
\par
The column ``excluding handedness'' means the number of bits after we have
taken into account that there is no true physical difference between a
model and the parity reflected model. If the number of bits in the two
columns do not deviate by the usual one bit, it is because
most of the models involved in the counting respect
parity invariance so
the counting is not changed by counting parity conjugate only for one.
\par
It is remarkable \cite{FroggattNielsen}
that imposing these four rather reasonable
assumptions: ``mass protection'', ``no anomalies'', ``charge quantization'',
and ``small representations'', we end up with only one model with
15 representations, namely the Standard Model itself.
Really, one can only construct
models with these assumptions in which the Weyl fermions occur in
``generations'' of the same irreducible representations as in
the Standard Model (listed in table 1).
Since we then have
a number of
irreducible representations which is divisible by 5
the assumptions above even explain approximately two
bits of the information listed under ``spin distribution''.
\par
In the column ``After understanding Weyl representations'' we have in
table 3 presented the numbers of bits left unexplained after
the understanding in terms of the four assumptions above about these
Weyl representations.

\subsection{Extracting information from the gauge group -
Why is the gauge group so ``skew'' ?}

A relatively {\em unbiased} way
of storing the
information of the gauge {\em group}
$ SMG = S(U_2 \times U_3) $, the 8 bits, is an observation
connected to the quantity $\chi= (\log q)/ r$, which we
have defined \cite{skewness} for all compact (potential gauge)
{\em groups}. For the Standard Model group ($SMG$)
the quantity
$\chi(SMG)=(\log 6) / 4$
is larger than for any other group, except that it takes just the
same value for the cartesian products $SMG \times SMG \times ... \times SMG$
of the standard model group
a number of times with itself. The quantity $q$ here is roughly associated
with the charge quantization rule(s) deduced from the group in question
and counts how many times smaller the quanta for the Abelian
charge(s) you can get by allowing the particles to couple to the semi-simple
part of the gauge {\em group}  than by not allowing that.
The information about physics to extract from the remarkable largeness
of this $\chi$ in Natures choice may stand as a bit of a challenge
to be found out, although some stability or robustness of a group
with high $\chi$ may be dreamt in a gauge theory which is speculated
to have quenched randomness like an amorphous medium.

A somewhat related property to this high-$\chi$-property is the property
that the Standard model {\em group} is remarkable by its relatively
low number of {\em automorphisms}
and ``generalized automorphisms'' (somewhat a concept invented for the
purpose; essentially some simple homomorphism) !
This principle of ``skewness'' \cite{skewness}
can - a bit favorably interpreted - also
be said to single out the Standard Model group as the ``most skew''
of the groups up to dimension 19. (I.e. taking into account groups
containing up to 19 gauge bosons).

With our convention of not counting the bigger
groups the principle of skewness should only be required to single out the
Standard Model group among groups up to dimension 12 (in fact,
it singles out the Standard Model group up to dimension 19)
and thus we can say that it explains the 8 bits for the groups
and replace it by ``the principle of skewness'' together with a couple of bits
telling the details of defining the concept of ``generalized automorphisms''
and about how to include in counting the infinitely many inner automorphisms
in order to really make the SMG win the competition of being skewest.

\subsection{Are the generation gaps a signal of
underlying approximative symmetries?}

A most striking feature of the quark and lepton masses is that they are
very small  - by a few orders of magnitude - compared to the
scale of masses $\sim 100 \; GeV$
(expected from the Standard Model)
coming from the
Fermi-constant. {Moreover}, the scale of
the masses for the three generations
of quarks and leptons differs (also) by orders of magnitude. These mass
gaps and the smallness of the masses
{may} suggest that there shall
exist at some level beyond the Standard Model - some
approximately {\em conserved}
quantum numbers (i.e. symmetries)
being different for right and left handed components
of the fermions in a way varying from generation to generation.
We have recently looked for schemes
\cite{CollinHolgerLowe}
in which
such quantum numbers are {\em gauged}.
{Something in the direction of}
repeating the Standard Model group for each generation could be helpful
(although the fact that the top quark mass is
much higher than the $\tau$ and the
b masses constitutes a problem). At least it is suggested
\cite{CollinHolgerLowe} that
some new quantum numbers exist which take on different values for different
generations.

\subsection{Why do we live in 3+1 dimensions ?
(only $\sim 3$ bits of information)}

There are several attempts to explain why we should have 3 spatial dimensions
(c.f. \cite{BarrowTipler} and references therein).
{In particular,
emphasis have been made as concerns the }
lack of stability of {motion in a Coulomb - or Newtonian - potential,}
if the space dimension were higher than three.
Mostly a single time dimension is
presupposed rather than being explained, but Jeff Greensite \cite{Greensite}
and also ourselves \cite{HolgerSvend1,HolgerSvend2}
seek to attack that bit of information
which tells about the signature of the metric of space time.
We (ourselves)  point out that just for the Weyl-equation in four space
time dimensions all linearly independent matrices  - $n^2$ of them,
when the Weyl field has $n$ components - are used
{\em just once each} in the
Weyl-equation\footnote{Strictly speaking the Weyl equation is not
defined in odd dimensions.  In that case we therefore just talk about what
should rather be called the Dirac equation.}: That
is to say that the requirement
\begin{equation}
n^2=\left\{ \begin{array}{ll}2^{d-2} & \mbox{for d even}\\
2^{d-1}& \mbox{for d odd} \end{array}\right\}=d
\end{equation}
apart from a trivial solution $n=d=1$ implies that the dimension
is $d=4$. Requiring the Weyl equation to come from a {\em real} action
leads then to the signature $3+1$ or $1+3$.
\par
As pointed out to us - recently - by Dharam Ahluwalia
at Los Alamos, the
space dimensionality three is also {\em the} dimension in
which the number of rotation generators
equals the number of Lorentz boost generators
(= the number of space dimensions).

\subsection{The values of the coupling constants }

The definition of the finestructure constants
$(\alpha_1, \alpha_2, \alpha_3)$ contain a  factor
$4 \pi$ which depends on conventions: $\alpha = g^2/4 \pi$.
In order to tell whether a
finestructure constant is small or big we
{need to} compare {it} with something that is a constant of
the same type (so as to avoid to make notation dependent searches for
regularities). A proposal \cite{DonHolger}
for making a  comparison to what is again finestructure constants is
to use the {\em multicritical} couplings - obtained from the
study of lattice gauge theories (Monte Carlo
calculations) - for the same group, and to avoid the
problem of the dependence on the scale $\mu$ of the ``running'' couplings we
compare the coupling constants at the Planck scale\footnote{Note,
that the effective gravitational coupling constant $\sim G \mu^2$
runs with the scale $\mu$ for trivial reasons, since $G$ has
dimension $m^{-2}$, and is -
by definition of the Planck scale
- of order $1$ just at the Planck scale.
If in some sense
the gravitational coupling constant would be ``multicritical'', it
would be at the Planck scale, if we assume that the critical value for
$G \mu^2$ is of order $\sim 1$.}
\begin{equation}  \label{sixthreethree}
\frac{\alpha_{U(1)crit}}{\alpha_1(\mu_{Planck})}=6,\mbox{  }
\frac{\alpha_{SU(2)multicr.}}{\alpha_2(\mu_{Planck})}=3,\mbox{
}\frac{\alpha_{SU(3)multicr.}}{\alpha_3(\mu_{Planck})}=3.
\end{equation}
The non-Abelian couplings yield the two numbers 3 with such an accuracy
that about 1.5 bits tell that they are integers.
A toy model
has been constructed
\cite{DonHolger} which involves
the group $G=SMG\times SMG \times SMG$ at the
Planck scale and which arrives at the relations (\ref{sixthreethree})
in a natural way.

We remark\footnote{We thank Dr. Lawrence Hall concerning this point.}
that in supersymmetric $SU(5)$-G.U.T.
one also have that the value of the  $SU(5)$ coupling
constant, at the G.U.T. or Planck scale, is rather near the multicritical
value,
\begin{equation}
\alpha_5(\mu_{Planck})\mbox{  or  } \alpha_5(\mu_{GUT})
\approx \alpha_{SU(5)multicr.}
\end{equation}
and thus the principle of (multi)criticality of coupling constants
{appears to have}  at least two chances of ``hitting''.

Note, that there is only little ``order of magnitude'' surprise in
the values of the coupling constants. That is,
whereas the quark masses, or the strong angle $\Theta_{QCD}$, say, are
surprisingly small the values of the
coupling constants $\alpha_1, \alpha_2,
\alpha_3, \alpha_G \sim 1 $ are all of the expected ``order $\sim 1$''
at the Planck scale
provided one takes into account that the natural unit of a finestructure
constant is of the order $\sim 1/20$ (which are the critical or multicritical
values).

\section{Some final remarks}

Briefly stated, we have proposed a
way to count in bits
the information content in various parts of the Standard Model
(its structure and parameters). Our counting may be useful
in evaluating the ``explanatory power'' (in terms of bits) of
attempts to guess physics beyond the Standard Model:
A fundamental model has to explain more bits than the
number of bits which are required to define it.

We have considered some possibilities of
going beyond the Standard Model - using
the unexplained structure and parameters of the Standard Model
as the only source of
``experimental inspiration''.


It is, at present, hard to get
experimental data which {\em directly} probes how Nature operates at energies
substantially above $\sim 100 \; GeV$ (per particle) - up to which
energy scales the
Standard Model (of the electroweak and strong interactions)
operates so successfully.

It is hard to create such experimental data here on
earth.\footnote{Note, that cosmic radiation which
hits earth (here Nature itself --- most likely pulsars? ---
functions as an accelerator) contains particles which
are very energetic (up to $\sim 10^{10} GeV$ per particle has been observed).
The possibility of using this radiation for doing particle physics
is strongly limited by the low luminosity of the high energy
part of the spectrum.}
Moreover, cosmological data are up to the problem that only little
information survives - when the universe reach thermal equilibrium
phases (this appears to be the case in many stages in the evolution of
the universe!).
In each local region of space where thermodynamic
equilibrium is reached, basically only conserved quantum numbers survive,
like energy, baryon and lepton numbers and the separate types
of isotopes (if it is cold enough). All other information
reach the ``heat death'', cf. sec. 2.4.

If we can {\em not} obtain {\em direct} (i.e. supported by
experiments) information about how Nature
operates beyond $\sim 100 \; GeV $ we are thus forced to attempt to guess
- { from purely theoretical considerations} -
{ about how Nature} {\em might be}
{ constructed at such higher energies}
and we have to base such guesses solely on the information
contained in the parameters and the
structure of the Standard Model itself.

{It appears}, that it is important to {read as
closely as possible ``what is to be read'' in } the Standard Model
- and try to identify paradoxes and mysteries from which
to be inspired (in order to try to go beyond).

How much of physics at Planck scales, say, are we able to
predict from controlling physics at $\sim 100 \; GeV$ scales ?

{It is fair to say, that} there is no sufficiently simple
candidate today of a fundamental model which
offers an understanding of the numerical values of the
19 parameters and the structure of the Standard
Model. It is, however, logically possible that {not only one but}
{\em several} different models (some of them simple, some of them more
complicated, most of them very complicated?)
could be constructed which all have the Standard Model
as the infrared limit.\footnote{The viewpoint that many models operating in
the ``ultraviolet''
may lead to the same ``infrared'' physics
have been pursued in the project we call random dynamics,
cf. e.g. Froggatt \& Nielsen
\cite{FroggattNielsen} and references therein.}
and in that case it is not  possible  - lacking the guidance from
experiments performed in a dialogue with Nature -
to choose between the many possible fundamental
models.

However, it is also logically possible, that one particular
model is so extraordinary simple that we find it a miracle that it exists
{(and explain the Standard Model)}.
In this case we would be tempted to believe it.
(See, also, discussions in S.Weinberg \cite{Weinberg}).

A sufficiently simple principle explaining a miraculously large
number of parameters or features of the Standard Model could
also deserve to be believed.

In order to {be guided towards} such principles
we have {in this article} tried to
to read the Standard Model w.r.t. its
structure and parameters, in particular we have - as an exercise -
tried to measure quantitatively: {\em How much information is there
to be inspired from in the Standard Model?}

It is presumably impossible to measure the simplicity of a model or principle
unbiased, i.e. without theoretical prejudices.
Nevertheless we have attempted to measure simplicity in terms of bits and
found --- {with rather natural conventions} ---
the information content in the Standard Model
(parameter system and structure) to be of the order of $2 \times 10^2$ bits.
Bits were assigned to the knowledge of the 19 external parameters,
the gauge Lie algebra, its representation system, and (a few bits
to) the dimension of space time, $3+1$.
Some parts of the structure of the Standard Model is very difficult
to translate into ``bits'' of information --- How many ``bits'' of information
is there in the principle of Lorentz invariance? ---
Thus, somewhat
arbitrarily, we stopped  before assigning bits to the principles
of Lorentz invariance
and quantum mechanics. It is not sufficiently
clear which alternatives to these principles we should consider.
We have also not counted the principle of gauge symmetry,
but indirectly some bits related to it were
included in assigning bits to the number of particles with different
spin and {\em helicity}.

Despite the arbitrariness of the measure we believe that the number
contains some
truth about how close theoretical physics is to catching up with experiment.

The $\sim 2 \times 10^2$ of unexplained information in the Standard Model
is very little as compared to the amounts of
unexplained information in other periods of the history of physics.
For example the large amounts of hadronic resonances and scattering data
which were collected in the sixties, say,
make up much more information than
$2 \times 10^2$ bits - but are now believed to be
{\em in principle} understood in QCD.

With this limited amount of unexplained experimental data available it is
necessary to use the precious bits with care and make sure, when proposing a
model, that it explains a positive amount of information.
We have made several attempts, some are reported in sec.3.2-3.6,
to find principles that could explain some bits of the
unexplained information in the Standard Model.
{Most of these principles or explanations,
however, cost more bits than they explain}.

Of the $\sim 3 \times 10^2$ bits of information
in the Standard Model, which we find in
a first calculation (cf. section 2.2),
the $\sim 90$ bits from the Weyl representations and a couple of bits from the
information on the numbers of particles are explained by four rather
reasonable assumptions  ``mass protection'', ``no anomalies'',
``charge quantization''
and ``small representations'' and gets compactified to only two bits
by describing the gauge group instead of the Lie algebra. Thereby we get down
to {the already mentioned}
$\sim 2 \times 10^2$ bits.

\begin{table}
\begin{tabular}{||c|c|c||} \hline
\multicolumn{3}{|c|}{Standard Model information}\\ \hline
&
\begin{tabular}{c}
Before understanding \\
Weyl representations \\
\end{tabular} &
\begin{tabular}{c}
After understanding \\
Weyl representations \\
\end{tabular} \\
\hline
Real parameters     & 155 & 155   \\ \hline
Discrete parameters & 115 &  21   \\ \hline
Total               & 270 & 176   \\ \hline
\end{tabular}
\caption{Resume of our counting of the measured but unexplained information
in the Standard Model. The first column does not utilize the explanation given
in section 3.2}
\end{table}

If we accept the explanation \cite{skewness}
of the resulting gauge group as
the winner in the game of the largest $\chi$ we convert the 8 bits of the
group into whatever we take the $\chi$-concept to ``cost'' (that could though
easily be more than 8 so that it would not pay). The finestructure constants
from multicriticality \cite{DonHolger}
at first glance explains of the order of 15 bits (three
numbers with 7\% accuracy), but again the price may be higher, since
these predictions require the rather complicated group
$SMG\times SMG \times SMG$ at the fundamental (Planck) level.
Unless we,
somehow, argue that a cross product of a single group
with itself several times is especially simple
the cost in bits would be appreciably
bigger than three times that for $SMG$ itself, i.e. $3\cdot 8$=24, so it would
not pay. Again the attempt \cite{CollinHolgerLowe}
to fit the quark and lepton mass spectra
does not pay;
it starts out with a rather low
reduction in unexplained information,
since it only explains the fact that there are
hierarchies but it does not predict
any detailed numbers - except after
fitting lots of parameters.\footnote{If we restrict attention to the
{\em principle} itself (without detailed models)
of having extra mass protecting quantum numbers
that idea may be so simple that even the tiny amount of
fitting explanation and the explanation of the big mass ratios is
indeed enough to give a positive gain!}
Hall et al. \cite{Hall}
is more promising starting out by explaining both
couplings, masses and mixing angles (constituting an information
content around $\sim 30$ bits.
Even if only of the order of four are left - arrived at by subtraction of
the information content in the input structure -
this is quite impressive already).
If we could combine with the bits by claiming the GUT SU(5) coupling
multicritical, we might gain two bits, say, minus again the price for
stating multicriticality (which may though hardly pay
for only one parameter).

More generally,
in seeking the inspiration, we have pointed towards the more surprising
features such as (1) the genuine paradoxes and (2) surprising parameter
values on grounds of dimensional
arguments (cf. section 3).

Both from the finestructure constant story with the group
$SMG\times SMG\times SMG$ and from the generation gaps there appears
the suggestion that different generations must have
different  quantum numbers at the more fundamental level.  \\

\noindent
{\bf Are there any signals appearing from several features of the Standard
Model?}

A very important signal is:
Standard Model is just the {\em low energy tail} of an underlying
theory!

All particles we know are very special in the sense that they have no
fundamental mass terms in the Lagrangian. They are all mass-protected
by the same gauge symmetry - and can only get non-zero masses
through the Weinberg-Salam Higgs mechanism.
This mass protection is seen from the following two features
of the Standard Model:
\begin{enumerate}
\item The ``mass protection'' (cf. section 3.2) observed concerning the Weyl
representations: There is no occurence of a representation and its
charge conjugate.

\item Of the 13 bosons of the Standard Model the 12 of them are
gauge bosons (mass protected) and one of them is a Higgs particle.
\end{enumerate}

\noindent
In fact, we have only seen mass-protected particles! (the Higgs particle
has not been observed yet)
Why do we not see particles which have a fundamental mass in the
Lagrangian?
When Nature has this possibility to
construct a massive particle why should it then only use
the Higgs-mechanism?
A natural explanation is that Nature indeed uses this possibility of
giving particles fundamental masses, but such masses becomes so
large that we have not yet seen them.\footnote{E.g., the most
natural explanation for
only seeing those Weyl representations that describe massless particles is that
the other representations (pairs which are
the charge conjugate of each other, i.e
Dirac representations) also exists but their particles are too heavy to be
seen.
Note, that this also constitutes a sort of explanation of parity violation:
If the weak interactions were parity invariant, then mass terms for the
fermions would be allowed,
and since we only look at the low energy tail of the theory we would
not see them.}

Among the particles with fundamental masses there are presumably both
scalar fields, vector fields (?), fermions etc.
To a scalar field with a fundamental mass there can be associated
a group of ``mass protected'' particles which
then get masses of the order of magnitude of the mass of the
scalar field
(functioning as a ``Higgs'' field). All the particles in the
Standard Model, i.e. all the particles we have seen until now,
belong to only one such group. They are all associated to only
one scalar field. This suggests that we are looking at a very
isolated group of particles on the mass axis. This group
simply is the group
associated with the lightest scalar field (one of them has to
be the lightest!).

All this suggests that the Standard Model is the low energy tail of a
more fundamental theory.



Pointing in the same direction is
the non-perturbative inconsistency \cite{Rakow} of QED and thus
presumably the
Standard Model. This enforces the existence of a
physically existing cut-off, i.e.
the Standard Model {\em has} to be regulated by some other more fundamental
theory at some higher energy scale.

\section*{Acknowledgements}

Support from the Danish Natural Science Research Council  (Grant No. 11-8705-1)
is gratefully acknowledged. S.E.R. would like to thank Prof. John Negele and
Prof. Ken Johnson for hospitality extended to him at the Center for Theoretical
Physics. We all have pleasure in thanking for (mainly) travel-support the
EEC-grant CSI-D430-C. Hospitality at Berkeley, Los Alamos and Stony Brook is
also acknowledged. Much work was done during travels with
$\bigtriangleup$-airlines across the USA.

\end{document}